\documentclass[aps,prb,twocolumn,superscriptaddress]{revtex4-1}

\usepackage[T1]{fontenc}
\usepackage{epsfig}
\usepackage{amssymb} %additional symbol sets
\usepackage{amsfonts} %additional fonts, useful for mathematical symbols
\usepackage{mathtools} %improves equation appearance
\usepackage{dcolumn} %makes columns in tables align by decimal point
\usepackage{graphicx} %allows inclusion of graphics
\usepackage{color}  %introduces colors ("monochrome" feature useful for BW preview)
\usepackage{bm}  %allows bolding of mathematical symbols
\usepackage{comment} %allows large blocks of commenting
\usepackage{units}%allows \nicefrac to be used
\usepackage{sidecap} %allows side captions to be used
\usepackage{textcomp} %gives a \textdegree command
\usepackage{units}%allows \nicefrac to be used
\usepackage{sidecap} %allows side captions to be used
\usepackage{textcomp} %gives a \textdegree command
\usepackage{mathrsfs}
\usepackage{enumitem} % allows compact lists
\usepackage{xfrac}

\usepackage{array} %allows text in table to be vertically centered
\usepackage{upgreek}
\usepackage{tabularx}
\usepackage{latexsym}
\usepackage{bm,array}
\usepackage{amsfonts}
\usepackage{amssymb}
\usepackage{amsmath}
\usepackage{bigstrut}
\usepackage{placeins}
\usepackage{siunitx}

\newcommand{\tref}[1] {Table~\ref{#1}}

\newcommand{\fref}[1] {Fig.~\ref{#1}}

\newcommand{\eref}[1] {Eq.~\ref{#1}}

\newcommand{\bpm}{\begin{pmatrix}}
\newcommand{\epm}{\end{pmatrix}}
\newcommand{\be}{\begin{eqnarray}}
\newcommand{\ee}{\end{eqnarray}}
\newcommand{\ba}{\begin{array}}
\newcommand{\ea}{\end{array}}

\def \yms{YbMnSb$_2$}

\begin{document}
\title{Observation of two-dimensional Fermi surface and Dirac dispersion in \yms}
%\title{Comparison of the magnetic properties of \blio\, and \glio\, harmonic honeycomb iridates}

\author{Robert Kealhofer}
\affiliation{Department of Physics, University of California, Berkeley, California 94720, USA}
\affiliation{Materials Sciences Division, Lawrence Berkeley National Laboratory, Berkeley, California 94720, USA}

\author{Sooyoung Jang}
\affiliation{Department of Physics, University of California, Berkeley, California 94720, USA}
\affiliation{Materials Sciences Division, Lawrence Berkeley National Laboratory, Berkeley, California 94720, USA}
\affiliation{Advanced Light Source, Lawrence Berkeley National Laboratory, Berkeley, California 94720, USA}

\author{Sin\'ead M. Griffin}
\affiliation{Department of Physics, University of California, Berkeley, California 94720, USA}
\affiliation{Molecular Foundry, Lawrence Berkeley National Laboratory, Berkeley, California 94720, USA}

\author{Caolan John}
\affiliation{Department of Physics, University of California, Berkeley, California 94720, USA}
\affiliation{Materials Sciences Division, Lawrence Berkeley National Laboratory, Berkeley, California 94720, USA}

\author{Katherine A. Benavides}
\affiliation{Department of Chemistry and Biochemistry, The University of Texas at Dallas, Richardson, Texas 75080, USA}

\author{Spencer Doyle}
\affiliation{Department of Physics, University of California, Berkeley, California 94720, USA}
\affiliation{Materials Sciences Division, Lawrence Berkeley National Laboratory, Berkeley, California 94720, USA}

\author{T. Helm}
\affiliation{Max Planck Institute for Chemical Physics of Solids, N\"othnitzer Stra\ss e 40, 01187 Dresden, Germany}

\author{Philip J. W. Moll}
\affiliation{Max Planck Institute for Chemical Physics of Solids, N\"othnitzer Stra\ss e 40, 01187 Dresden, Germany}

\author{Jeffrey B. Neaton}
\affiliation{Department of Physics, University of California, Berkeley, California 94720, USA}
\affiliation{Molecular Foundry, Lawrence Berkeley National Laboratory, Berkeley, California 94720, USA}

\author{Julia Y. Chan}
\affiliation{Department of Chemistry and Biochemistry, The University of Texas at Dallas, Richardson, Texas 75080, USA}

\author{J. D. Denlinger}
\affiliation{Advanced Light Source, Lawrence Berkeley National Laboratory, Berkeley, California 94720, USA}

\author{James G. Analytis}
\affiliation{Department of Physics, University of California, Berkeley, California 94720, USA}
\affiliation{Materials Sciences Division, Lawrence Berkeley National Laboratory, Berkeley, California 94720, USA}

\date{\today}

\begin{abstract}
We present the crystal structure, electronic structure, and transport properties of the material \yms, a candidate system for the investigation of Dirac physics in the presence of magnetic order.  Our measurements reveal that this system is a low-carrier-density semimetal with a 2D Fermi surface arising from a Dirac dispersion, consistent with the predictions of density functional theory calculations of the antiferromagnetic system. The low temperature resistivity is very large, suggesting scattering in this system is highly efficient at dissipating momentum despite its Dirac-like nature.
\end{abstract}

%\pacs{61.05.C,65.40.Ba,71.15.Ap,71.15.Mb,71.18.+y,79.60.Bm}

\maketitle

\section{Introduction}
The coexistence of topological band structures and more conventional broken symmetry orders has yet to be investigated extensively, primarily due to the lack of suitable materials systems.  As a result, several studies have focused on the structural family \textit{R}Mn\textit{Pn}$_2$, where \textit{R} is a rare-earth metal and \textit{Pn} is a pnictide (usually Sb or Bi),\cite{wang_multiband_2012, guo_coupling_2014, jo_valley-polarized_2014,may_effect_2014,farhan_aemnsb_2014,liu_discovery_2015} due to the possible coexistence of protected band crossings and magnetic order on the Mn sublattice.  The magnetic order is thought to be antiferromagnetic, breaking the spin degeneracy of the system, and it has been argued that this leads to crossings separated in energy in the band structure.\cite{zyuzin_weyl_2012,borisenko_time-reversal_2015,wang_time-reversal-breaking_2016}  Two members of this family, SrMnBi$_2$ and YbMnBi$_2$, have been proposed as possible topological semimetals hosting magnetic order,\cite{jo_valley-polarized_2014,wang_magnetotransport_2016} but comparatively little work has investigated their Sb-based cousins.\cite{huang_nontrivial_2017}  We have synthesized a new member of this family, \yms, and report here comprehensive transport and spectroscopic measurements. Evidence from Shubnikov-de Haas oscillations and angle-resolved photoemission spectroscopy (ARPES) suggest there is at least one 2D Fermi surface of Dirac origin.  The ARPES data is in agreement with density functional theory (DFT) calculations of the antiferromagnetic band structure, which indicates that \yms\ may therefore be a new topological material in the presence of magnetic order. However, despite the observation of quantum oscillations, the low temperature resistivity of these materials is large, suggesting that quantum and transport lifetimes have the same origin in these materials.

\section{EXPERIMENT}
\subsection{Crystal synthesis and structure}
Single crystals of \yms\ were synthesized using a tin-flux technique.  Ytterbium (99.9\%), manganese (99.95\%), antimony (99.9999\%), and tin (99.999\%), all from Alfa Aesar, were mixed in the mole ratio 1:1:4:10 (Yb:Mn:Sb:Sn).  The mixture was placed in an alumina crucible and sealed in an evacuated quartz ampule, then heated over 8\,h to 1050\textdegree C where it dwelled for 24\,h.  Next, the ampule was cooled to 600\textdegree C over 100\,h, and then centrifuged to remove excess tin.  This process yielded single crystals of approximately 1\ mm $\times$ 1\ mm $\times$ 0.05 mm.

\begin{figure}
    	{\includegraphics[width=8.6cm]{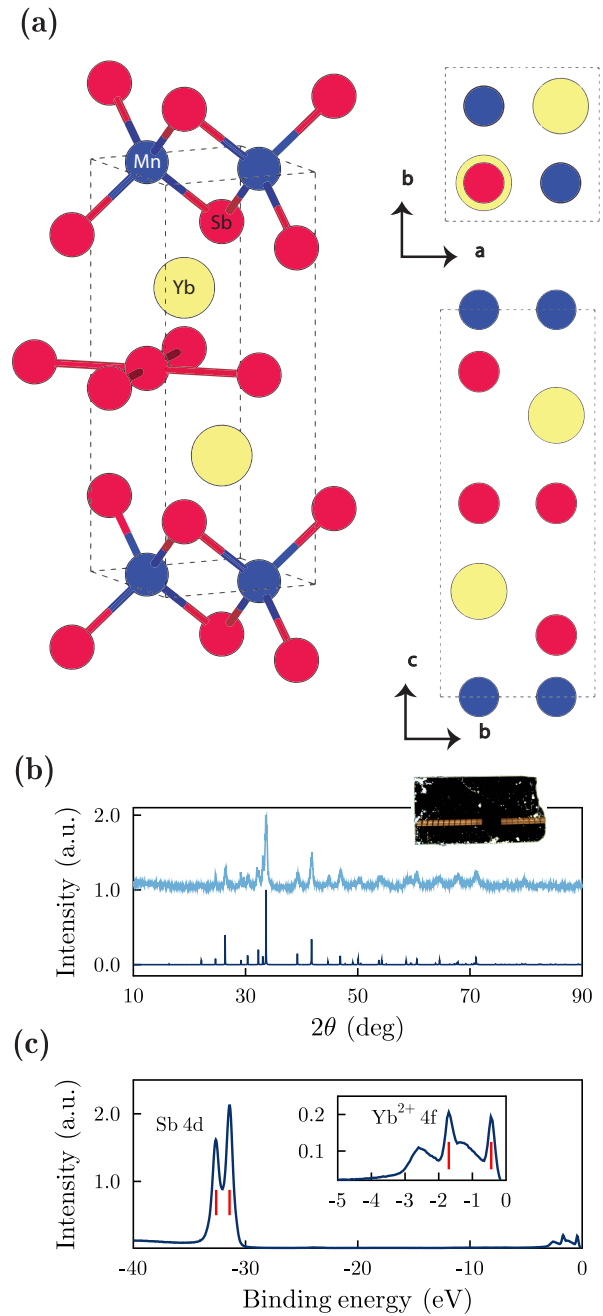}
  	\caption{The structure of \yms.  
(a) Schematic of the \yms\ structure drawn in {\scshape vesta}, from single crystal X-ray diffraction data.  \yms\ is isotypic with the HfCuSi$_2$ structure type.  Here yellow is Yb, blue is Mn, and red is Sb.  Perspective drawing (left) shows Sb planes and the tetrahedral coordination of Mn by Sb, top right looks down the $c$-axis, and bottom right looks down the $a$-axis.  
(b) Experimental powder pattern of \yms\ crystals (top trace) and expected powder pattern from \tref{struct_param} and \tref{atom_pos}, simulated in {\scshape vesta}.  Inset: Millimeter-sized \yms\ crystal.  Crystals are highly lustrous, as can be seen from the reflection of the fluorescent room lighting across the bottom of the crystal.
(c) Core level spectroscopy of \yms\ unambiguously indicates the 2+ oxidation state of Yb. % Red lines indicate core level peaks associated with each labeled species.
}
  	\label{fig1}}
\end{figure}

The crystallographic parameters are listed in \tref{struct_param} and the atomic coordinates and displacement parameters are provided in \tref{atom_pos}.  The crystal structure is shown schematically in \fref{fig1}(a). Single crystals of \yms\ were cut to an appropriate size and mounted on a glass fiber using epoxy.  The fiber was mounted on a Bruker D8 Quest Kappa single-crystal X-ray diffractometer with a Mo K$\alpha$ I$\upmu$S microfocus source ($\uplambda = 0.71073$\AA) operating at 50\,kV and 1\,mA, a HELIOS optics monochromator, and a CMOS detector.  The Bruker program {\scshape sadabs} (multi-scan method) was used to correct the collected data for absorption.  A starting model of \yms\ was obtained using the intrinsic phasing method in {\scshape shelxt}.\cite{sheldrick_shelxt_2015}  Atomic sites were refined anisotropically using {\scshape shelxl2014}.\cite{sheldrick_crystal_2015}  Figure~1(b) shows a characteristic powder pattern (collected using Cu $K\alpha$ radiation on a Rigaku Ultima IV powder diffractometer) together with the simulated powder pattern, computed in {\scshape vesta}.\cite{momma_vesta_2011} Core-level spectroscopy on the \textsc{merlin} beamline (see Sec. II-C) confirms the divalent oxidation state of Yb (\fref{fig1}(c)).

\yms\ belongs to the tetragonal, centrosymmetric \textit{P}4\textit{/nmm} space group and is isotypic with the HfCuSi$_2$ structure type.\cite{andrukhiv_structure_1975}  \yms\ does not exhibit a Mn deficiency that is observed in compounds of the same structure type with larger rare earth elements, e.g. LnMn$_{1-x}$Sb$_2$ (Ln = La, Ce, Pr, Nd, or Sm).\cite{wollesen_ternary_1996}  The independent confirmation of the \yms\ crystal structure allows for a first-principles calculation of the electronic band structure, which we consider next.%Figure~1(c) confirms the divalent oxidation state of Yb.

%{\cblue Ver.1 ARPES measurements were performed at the MERLIN beamline 4.0.3 of the Advanced Light Source employing both linear horizontal and linear vertical polarization from an elliptically polarized undulator. A Scienta R8000 electron spectrometer with 2D parallel detection of electron kinetic energy and angle in combination with a highly-automated six-axis helium cryostat goniometer with 6 K base temperature operating in low 10$^{-11}$ torr pressure. Total energy resolution of approximately 10-20 meV was used for measurements.}

%{\cblue Ver.2 ARPES measurements were performed at the MERLIN beamline 4.0.3 of the Advanced Light Source employing 78 eV photons. Samples were cleaved $in-situ$ in an ultrahigh vacuum in low 10$^{-11}$ torr pressure. DFT calculation.....}

%Single crystals of \blio\ and\glio\ were synthesized using a vapor transport technique. Ir (99.9\% purity, BASF) and Li$_2$CO$_3$ (99.999 \% purity, Alfa-Aesar) powders were grinded and cool-press pelletized in the ratio of 1:1.05. The pellets were placed on an alumina crucible, reacted for 12hrs, and then cooled down to room temperature at 2 \textdegree C/hr to yield single crystals of either \blio\ or\glio. The distinction between the two growth is the dwelling temperature: reacting the pellet at 1,050\textdegree C yields single crystals of \blio, while 1,000\textdegree C gives off \glio.
%
%\begin{figure}[ht]
%    	{\includegraphics[width=9cm]{Figures/Fig3_hhc.pdf}
%  	\caption{(Color online) {\bf Crystal Structure of  $\beta$-Li$_2$IrO$_3$} }
%  	\label{fig:hhc}}
%	\end{figure}

%The single crystals of both polymorphs are clearly faceted, and around 105$\times$150$\times$300 $\upmu$m$^3$ in size. Both structures have a very similar diamond facet and are, therefore, difficult to identify by morphology along. To distinguish between these different growth conditions, we performed powder and single crystal x-ray diffraction. The details of these refinements are summerized on the Supplementary information. 

\begin{table}
\caption{\label{struct_param}Crystallographic parameters.}
\begin{ruledtabular}
\begin{tabular}{ll}
\textbf{Crystal type} & \textbf{\yms} \\
space group & \textit{P}4\textit{/nmm}, \#129\\
\textit{a, c} (\AA) & 4.3215(17), 10.828(4)\\
\textit{V} (\AA$^3$) & 202.22(17)\\
\textit{Z} & 2 \\
temperature (K) & 300\\
$\uptheta$ range (\textdegree) & 3.8--30.7\\
$\upmu$ (mm$^{-1}$) & 38.93\\
measured reflections & 1085\\
independent reflections & 224\\
$R_{\mathrm{int}}$ & 0.033\\
$\Delta\uprho_{\mathrm{max}}$ (e \AA$^{-3}$) & 1.64\\
$\Delta\uprho_{\mathrm{min}}$ (e \AA$^{-3}$) & -3.56\\
GOF & 1.27 \\
$R_1(F)\footnote{$R_1=\Sigma \mathbf{|}|F_0|-|F_c|\mathbf{|}/\Sigma|F_0|$}$ & 0.033\\
$wR_2\footnote{$wR_2 = \left\{\Sigma\left[w(F_0^2-F_c^2)^2\right]/\Sigma\left[w(F_0^2)^2\right]\right\}^{1/2}$}$ & 0.060\\
\end{tabular}
\end{ruledtabular}
\end{table}

\begin{table}
\caption{\label{atom_pos}Atomic positions.}
\begin{ruledtabular}
\begin{tabular}{llllll}
\textbf{\yms} & \textbf{Wyckoff site} & \textbf{x} & \textbf{y} & \textbf{z} & \textbf{$\mathbf{U_{\mathrm{eq}}}$ (\AA$^2$)} \\
Yb & 2\textit{c} & \sfrac{1}{4} & \sfrac{1}{4} & 0.72728(8) & 0.0106(2) \\
Mn & 2\textit{a} & \sfrac{3}{4} & \sfrac{1}{4} & 0 & 0.0115(6)\\
Sb 1 & 2\textit{b} & \sfrac{3}{4} & \sfrac{1}{4} & \sfrac{1}{2} & 0.0115(3)\\
Sb 2 & 2\textit{c} &  \sfrac{1}{4} & \sfrac{1}{4} & 0.15995(12) & 0.0099(3)\\
\end{tabular}
\end{ruledtabular}
\end{table}

%\begin{table}[!hb]
%\begin{center}
%\caption{Structural parameters at room temperature}
%{\renewcommand{\arraystretch}{1.5}
%\begin{tabularx}{0.45\textwidth}{ >{\setlength\hsize{1\hsize}\centering}c|>{\setlength\hsize{1\hsize}\centering}X >{\setlength\hsize{1\hsize}\centering}X }
%\hline \hline\
% & {\bf \yms} \tabularnewline 
%\hline
% Space group:  & \textit{P4/nmm} (\#129) \tabularnewline 
% Z: & 2 \tabularnewline 
%{\bm $a$,$c$ (\AA):} & 4.3215(17), 10.828(4) \tabularnewline 
%{\bm $\alpha$,$\beta$,$\gamma$ (\textdegree):} & 90, 90, 90 \tabularnewline 
%Volume (\AA$^3$): & 202.22 \tabularnewline
%\hline \hline 
%\end{tabularx}}
%\label{struct_param}
%\end{center}
%\end{table}

\subsection{Electronic structure}
\begin{figure}
    	{\includegraphics[width=9cm]{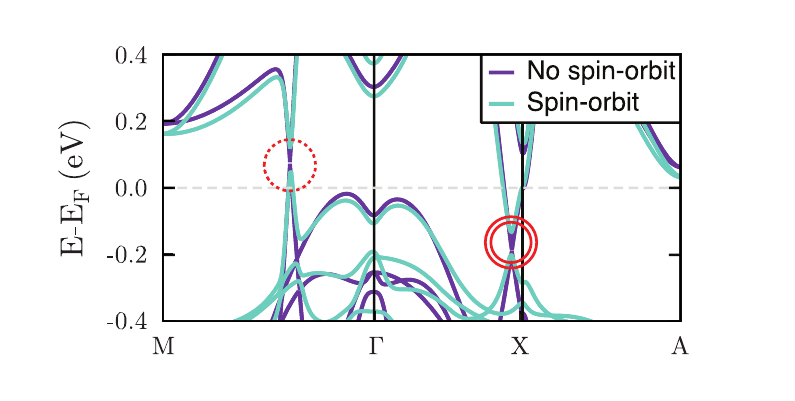}
  	\caption{Electronic structure of \yms. Note the band crossings on the $\Gamma$-M and $\Gamma$-X lines, indicated with a dashed circle and a double circle respectively.
}
  	\label{bandstruct}}
\end{figure}

Using the lattice parameters using the measured structural data reported in \tref{struct_param} and \tref{atom_pos}, we performed spin-polarized density functional calculations within the generalized gradient approximation (GGA) using the Vienna \textit{ab initio} Simulation Package ({\scshape vasp}).\cite{VASP1, VASP2} We used the projector augmented wave (PAW) method, treating explicitly 5\textit{s}, 5\textit{p}, 4\textit{f}, and 6\textit{s} for Yb; 3\textit{d} and 4\textit{s} for Mn; and 5\textit{s} and 5\textit{p} for Sb.  Wavefunctions were expanded in a plane wave basis up to an energy cutoff of 600\,eV.  We used Monkhorst-Pack $k$-point grids of $10\times10\times4$ for Brillouin zone sampling.\cite{Monkhorst/Pack:1976}  Spin-orbit coupling was neglected except when computing the band structure in \fref{bandstruct} and when identifying the magnetic easy axis.  We used the PBE GGA functional\cite{PBE1} with Hubbard $U$ corrections\cite{Dudarev_et_al:1998} for the Yb 4\textit{f} states, where we used a $U_{\mathrm{eff}}$ of 4\,eV.  This value of $U_{\mathrm{eff}}$ was chosen to best reproduce the photoemission data, which placed the Yb 4\textit{f} states at $-0.6$\,eV below the Fermi energy.  We did not include a Hubbard $U$ for the Mn 3\textit{d} electrons after tests indicated such a $U$ had no effect on the electronic structure near the Fermi level.

All calculations assumed antiferromagnetic (AFM) order for the spins on the Mn sites, with the two Mn ions in each cell having opposite spin.  This AFM ordering was calculated to be 0.23\,eV and 2.62\,eV per formula unit lower in energy than ferromagnetic order and a non-spin-polarized calculation, respectively.  We compute the magnetic easy axis to be in-plane, and 1\,meV/f.u. lower in energy than for out-of-plane spin orientations.

The band structure is shown in \fref{bandstruct}.  Note the linear band crossings on the $\Gamma$-M and $\Gamma$-X lines.  The calculated Fermi surface is displayed in \fref{fig4}, showing 2D and 3D Fermi surface pockets.

\subsection{Angle-resolved photoemission spectroscopy}
To experimentally investigate the electronic structure of \yms, we performed ARPES measurements at the MERLIN beamline 4.0.3 of the Advanced Light Source, employing both linear horizontal and linear vertical polarization from an elliptically polarized undulator.  The experiment used a Scienta R8000 electron spectrometer with 2D parallel detection of electron kinetic energy and angle in combination with a highly-automated six-axis helium cryostat goniometer with 6\,K base temperature operating in low 10$^{-11}$\,torr pressure. The energy resolution of these measurements was approximately 10--20\,meV.  These results are presented in \fref{fig4} and \fref{fig5}.

\subsection{DFT/ARPES comparison}
\begin{figure}
    	{\includegraphics[width=8cm]{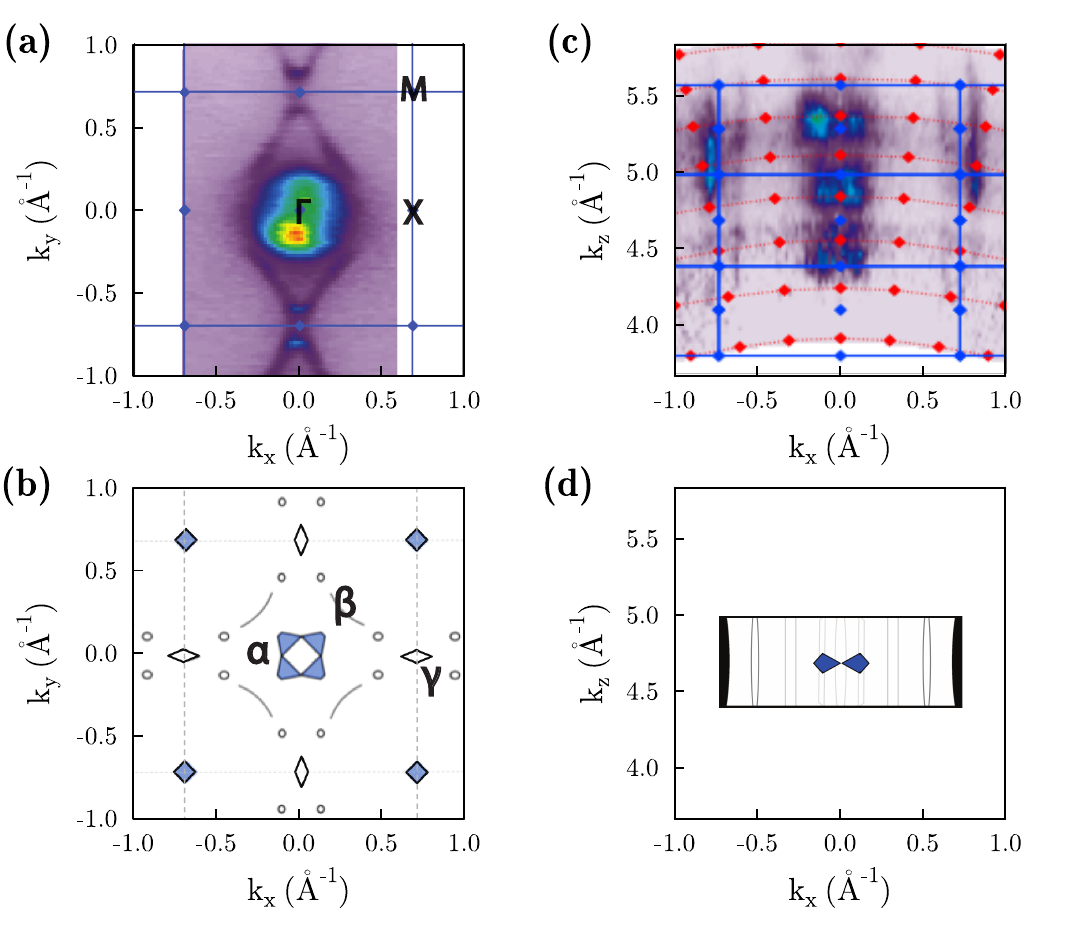}
  	\caption{ {
(a) Fermi surface (FS) map at 78\,eV photon energy. Yellow and red dashed lines indicate the high symmetry line for (c) and (e), respectively.
(b) Calculated FS with a 30\,meV energy resolution.   
(c) Photon energy dependence of the ARPES intensity plot.  
(d) $k_y=0$ cut of the FS in (b).  Features off the $k_y=0$ plane are shown in dashed lines.
%(e) ARPES intensity plot of \yms\ along the M-$\Gamma$-M line. The calculated band dispersions (black lines) are plotted for comparison.  The calculated dispersion is scaled by a factor of two to fit the data.
}
  	\label{fig4}}}
\end{figure}

Figure~3(a) shows a Fermi surface (FS) map obtained by integrating ARPES intensity inside a 20\,meV energy window around the Fermi energy ($E_F$), measured with photon energy 78\,eV. There are three distinct regions with high photoemission intensities: a 3D FS centered at the $\Gamma$ point ($\alpha$), a small 2D hole pocket along the $\Gamma$-X line ($\beta$), and a skinny, 2D banana-like pocket on the $\Gamma$-M line ($\gamma$). To compare the data with the calculation, we plot the Fermi surface as calculated by DFT in \fref{fig4}(b).  This calculation shows all three of these features.  The lack of fourfold symmetry in the data (\fref{fig4}(a)) as compared with the calculated FS (\fref{fig4}(b)) is due to selection rules arising from the photon polarization. %{$\cblue we may want to describe the origin of these features which can be estimated  by band calculation in Fig.5(e).} 

Figure~3(c) shows the photon-energy dependent $k$-maps from 50--120\,eV of the ARPES intensity at $E_F$, cut along the $\Gamma$-X high-symmetry line (yellow dashed line in \fref{fig4}(a)). The Fermi-edge $k_x$-$k_z$ photon-energy dependent maps show 2D behavior for the X-point pocket ($\beta$), but 3D $k_z$-dispersive behavior near $\Gamma$ point (pocket $\alpha$). These features are reported in related compounds: The 2D state was observed in YbMnBi$_2$\cite{borisenko_time-reversal_2015} which was attributed to the Yb-Bi plane, and 3D behavior has been reported in LaAgSb$_2$, and attributed to the Sb plane. \cite{16petrovic}

\begin{figure}
    	{\includegraphics[width=8cm]{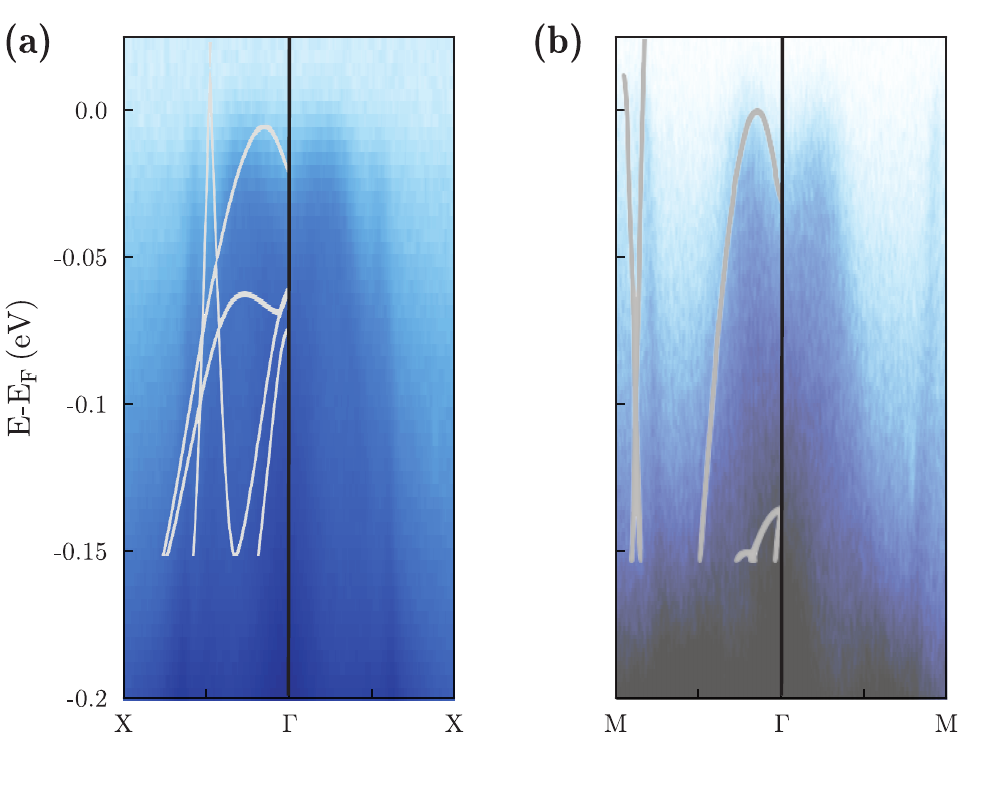}
  	\caption{ {ARPES intensity plot for (a) X-$\Gamma$-X and (b) M-$\Gamma$-M.  %DFT results are overlaid in grey for comparison.}
}
  	\label{fig5}}}
\end{figure}

ARPES intensity plots along the M-$\Gamma$-M and X-$\Gamma$-X lines are shown in \fref{fig5}. The calculated bands in each cut are overlaid on the ARPES data to show the robust agreement between the experiment and the calculation. Note the presence of a hole pocket in panel (a) and an electron pocket in panel (b), both with linear dispersion. 

%Calculations in Fig.~5(e) suggest that the intensity at the $\Gamma$ point (pocket $\alpha$) is related to two hole-like pockets originating from Sb 5$p_x$ and 5$p_y$ states. The banana-like FS on the $\Gamma$-M direction (pocket $\gamma$), on which we focus, is consistent with the calculated FS. The calculated FS near the X-point agrees well with calculated results for $(k_x,k_y) = (0,0.55)$\,\AA$^{-1}$, but we barely observed FSs near $(k_x,k_y) = (0.55,0)$\,\AA$^{-1}$. This discrepancy is most likely due to the polarization effect. Figure\,4(e) shows the ARPES intensity plotted for cuts along the $\Gamma$-M line as indicated in Fig.~4(a). The raw data and the corresponding calculated data depicted in Fig.~4(e) shows that the experimental band is almost linear at $\Gamma$-M/2 point with Dirac-like shape in good agreement with the theoretical calculation.

%The photon-energy dependent $k$-maps from 50 to 120 eV of the ARPES intensity at $E_F$ along the $\Gamma$-X high-symmetry line (yellow dashed line in Fig. 4(a), displayed in Fig. 4(c). Fermi-edge $k_x$-$k_z$ photon-energy dependent maps showing 2D surface state behavior for X-point, but 3D $k_z$-dispersive behavior near $\Gamma$ point. Those features are reported for different compounds: 2D surface state has been shown in YbMnBi$_2$ \cite{borisenko_time-reversal_2015} which is originated from Yb-Bi plane, 3D behavior has been reported for LaAgSb$_2$ compound which is related to Sb layer. \cite{16petrovic}

\subsection{Transport and Hall effect}
Single crystals were contacted by sputtering gold in the desired configuration and attaching annealed platinum wires with silver paste (Dupont 4929N).  After contacting, the sample was shaped into a ``Hall bar'' geometry using a plasma focused ion beam (FIB).  The resistance and magnetoresistance measurements were performed in a Quantum Design Physical Properties Measurement System (PPMS).

\begin{figure}
    	{\includegraphics[width=8cm]{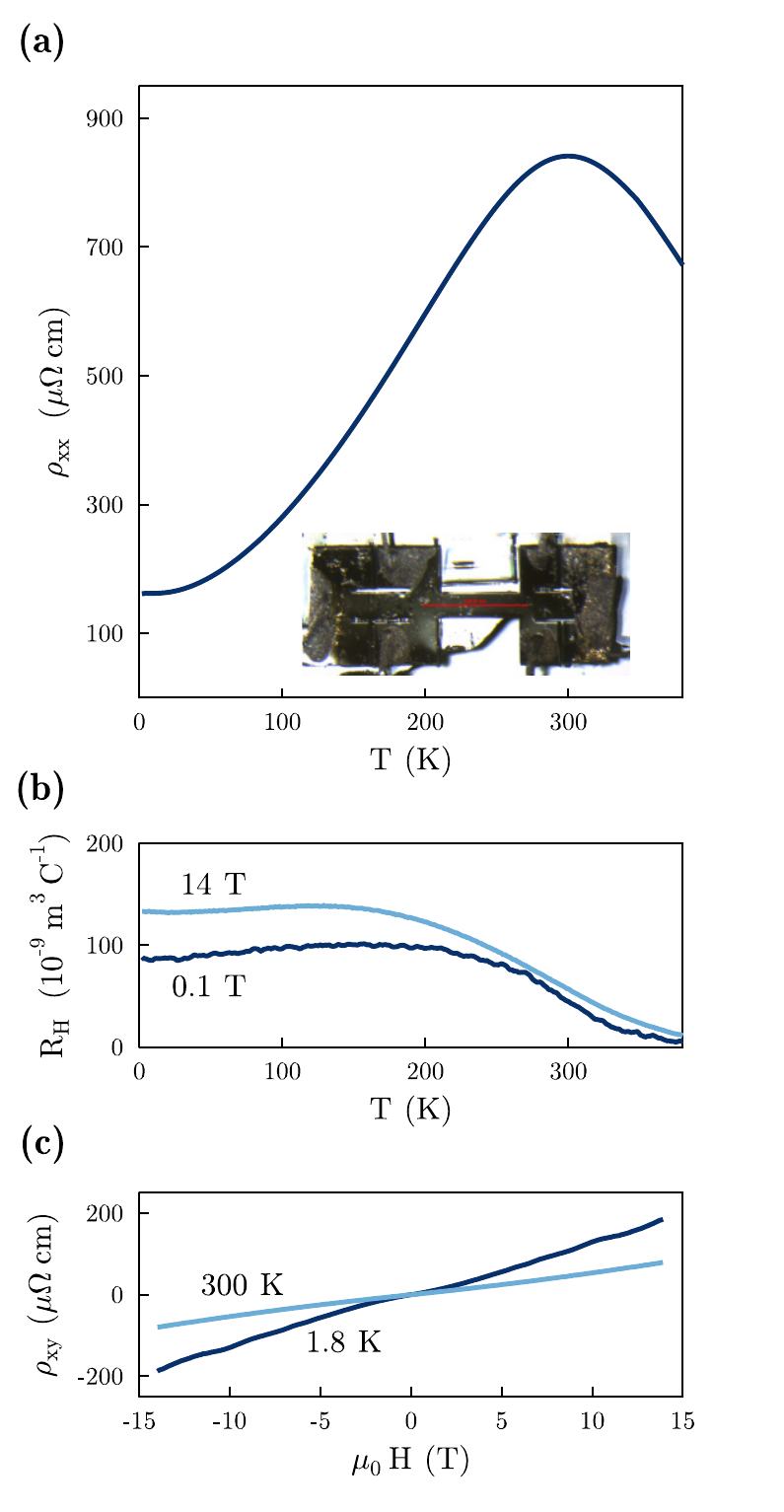}
  	\caption{ Magnetotransport and heat capacity of \yms.
(a) Resistivity versus temperature.  Inset: ``Hall bar'' sample geometry prepared using plasma FIB etching.
(b) Hall coefficient ($R_H$) versus temperature measured at $\pm0.1$\,T and $\pm14$\,T.  
(c) $\rho_{xy}$ versus magnetic field, antisymmetrized with respect to field polarity, at $1.8$\,K and 300\,K.
}
  	\label{fig2}}
\end{figure}

Figure~5(a) shows $\rho_{xx}$ as a function of temperature.  The non-monotonic temperature dependence of the resistivity could indicate a crossover between metallic and semiconducting behavior.  It is unclear that this peak is associated with a true phase transition as its position varies between samples by over 20\ K.  The plot of Hall coefficient (\fref{fig2}(b)) as a function of temperature was constructed by antisymmetrizing data at $\pm0.1$\,T and $\pm14$\,T.  The latter trace captures the high-field Hall coefficient, and the former trace approximately reflects the low-field Hall coefficient.  Above $T\gtrsim250$\,K, nonlinearity in the Hall effect has vanished and the two traces converge.

At high fields ($\omega_c \tau \gg 1$), the Hall coefficient of a multiband system is inversely related to the degree of compensation, according to
\begin{equation}
(e R_H)^{-1} \equiv n_{\infty} = n_h-n_e,
\label{hi_field}
\end{equation}
where $n_h$ is the total hole density and $n_e$ is the total electron density.\cite{kohler_magnetische_1949}  We observe a reduction of the Hall coefficient with increasing temperature, which could reflect a reduction in the degree of compensation (assuming the high field limit is reached by 14~T, at 1.8~K, $n_\infty\sim 4.7\times10^{19}$~cm$^{-3}$ while at 390~K $n_\infty\sim 7.5\times10^{20}$~cm$^{-3}$).  However, it is likely that the high field limit is not reached at high temperatures, so this change arises from a combination of the availability of thermally excited carriers and changing mobilities of electron and hole pockets.

A simple two-band low-field form of the effective carrier density predicts that this effective carrier density is dominated by high mobility carriers, according to
\begin{equation}
(e R_H)^{-1} \equiv n_0 = \frac{(n_h \mu_h + n_e \mu_e)^2}{n_h \mu_h^2 - n_e \mu_e^2},
\label{lo_field}
\end{equation}
where $\mu_i$ is each carrier's respective mobility. Band structure and photoemission both indicate that three bands may be present at the Fermi level, but qualitatively this two-band model shows that the Hall coefficient is determined by the highest mobility carrier.  At low temperature, then, the 0.1\,T trace (representing the low-field limit) shows the presence of high-mobility holes.

%\subsection{Heat capacity}
%Heat capacity was measured using a DC relaxation method.  The inset data in \fref{fig2}(a), plotted as $C_V/T$ against $T^2$, fall on a line, consistent with the textbook Sommerfeld and Debye theory. The reported Sommerfeld coefficient from the fit intercept is $-180\ \mu \mathrm{J\,mol}^{-1}\,\mathrm{K}^{-2}$ with a 95\% confidence interval of $(-1.5, 1.1)\ \mathrm{mJ\,mol}^{-1}\,\mathrm{K}^{-2} $.  The Debye temperature $\Theta_D$ can be extracted from the fit slope $\alpha$ according to
%\begin{equation}
%\label{th_d}
%\Theta_D = \left(R \frac{12 \pi^4}{5 \alpha}\right)^{1/3},
%\end{equation}
%where $R$ is the gas constant.  Here $\alpha = (1\pm0.05)$ $\mathrm{mJ~mol}^{-1}\ \mathrm{K}^{-4}$, corresponding to a Debye temperature of $(125\pm2)~\mathrm{K}$.  %Error bars represent a 95\% confidence interval.

\subsection{Shubnikov-de Haas measurements}
\begin{figure*}[ht]
    	{\includegraphics[width=17.2cm]{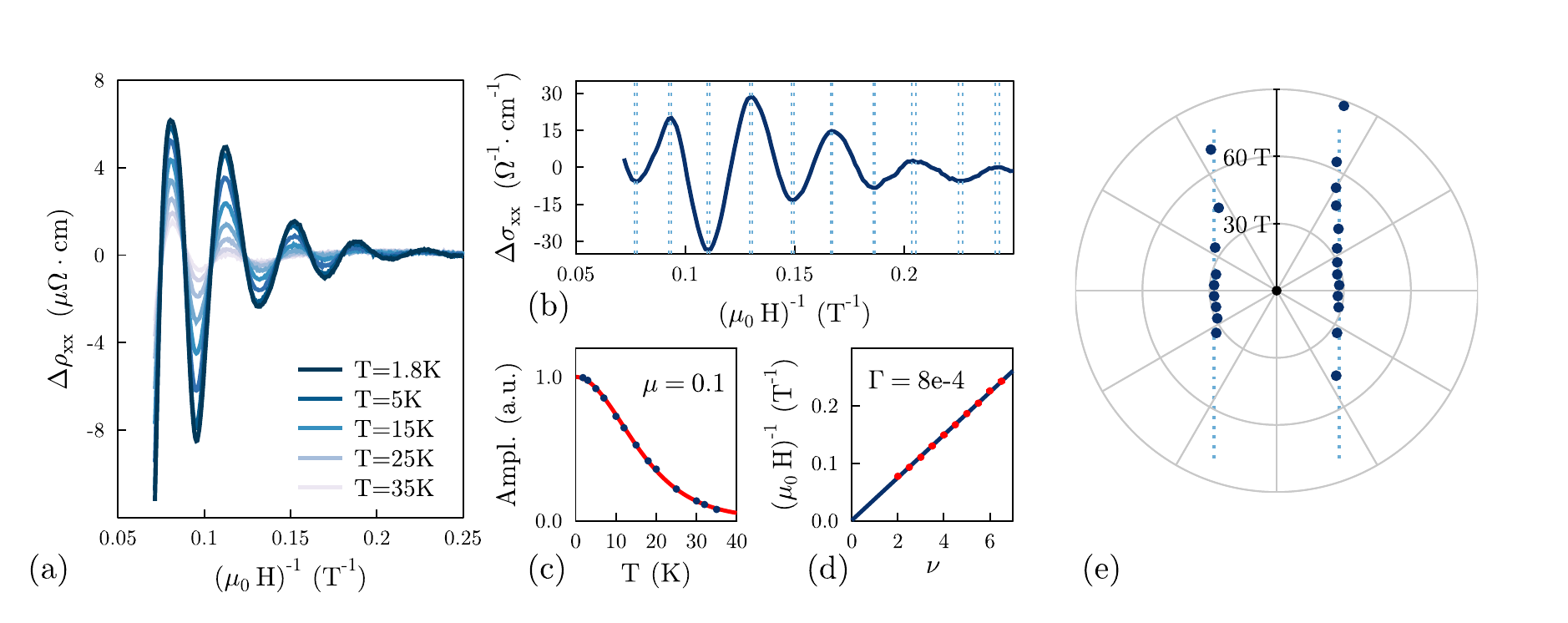}
  	\caption{ Shubnikov-de Haas oscillations in \yms.  See also \tref{qo}.
(a) Background-subtracted resistivity oscillations as a function of inverse field.  The oscillation frequency is $(28\pm1)$\ T, corresponding to an orbital area of $2.8 \times 10^{-3} $\ \AA$^{-2}$ (\eref{freq}).  
(b) Longitudinal conductivity, $\sigma_{xx}$, plotted against inverse field.  Dashed lines indicate the uncertainty in identifying the peak locations.  Minima in the conductivity mark integer Landau levels crossing the Fermi surface.  $\sigma_{xx}$ is computed using \eref{sxx}.
(c) Lifshitz-Kosevich (LK) (\eref{LK1} \& \eref{LK2}) fit to decay of the background-subtracted oscillation amplitude with temperature, with $\mu=0.100\pm0.001$.  Blue markers indicate oscillation amplitudes, and the red curve is the LK fit.  
(d) Landau fan diagram with linear fit.  Using \eref{LL}, the most likely $\phi_B$ is $(0.9984\pm0.003)\pi$ with the error bars computed from the fit error.  
(e) Polar plot of angular dependence of oscillation frequency.  Dark blue markers indicate oscillation frequencies at different angles, as determined via Fourier transform of the background-subtracted oscillations.  At zero degrees, the current flows (and voltage is measured) along the $a$-axis and the field is aligned with the $c$-axis.  At ninety degrees, the field and current are parallel to the $a$-axis.  The dispersion of the oscillation frequency with angle is consistent with a two-dimensional Fermi surface, which would disperse as $1/\mathrm{cos}(\theta)$ (dashed lines).  Above 75 degrees, the oscillation amplitude becomes too small to track.
  	\label{fig3}}}
\end{figure*}

We observed Shubnikov-de Haas (SdH) oscillations in magnetic fields of up to 14\,T at temperatures between 1.8\,K and 35\,K.  Our results (\fref{fig3}) indicate that the Fermi surface is small (perpendicular area approximately 0.1\% of the bulk Brillouin zone), strongly anisotropic, and the effective mass is small ($\mu=m^*/m_e=0.1$).

Raw resistance versus field traces are processed in several steps.  First, we subtract a best-fit fourth-order polynomial background from the trace.  (For the frequencies observed, the results are largely insensitive to the order of polynomial background subtracted.)  Our background-subtracted traces are plotted against inverse field ($1/\mu_0 H$) in \fref{fig3}(a).  The background-free data is Fourier transformed against the inverse field to confirm the oscillation frequency.  

In the absence of magnetic breakdown and other high-field effects, the oscillation frequencies observed are proportional to the areas of extremal orbits on the Fermi surface, in the plane normal to the applied field.  Explicitly,
\begin{equation}
F=\frac{\hbar}{2 \pi e} S_{e},
\label{freq}
\end{equation}
where $F$ is the oscillation frequency and $S_e$ the ($k$-space) extremal area.  

The SdH effect appears intrinsically as an oscillatory component of the conductivity, which we are able to obtain by simultaneous measurements of Hall and longitudinal resistance. Then $\sigma_{xx}$ can be calculated according to
\begin{equation}
\sigma_{xx}=\frac{\rho_{xx}}{\rho_{xx}^2+\rho_{xy}^2}
\label{sxx}
\end{equation}
and is shown in \fref{fig3}(b).

The SdH oscillation amplitude decays with temperature according to the Dingle form
\begin{equation}
R_T=\frac{x}{\mathrm{sinh}(x)},
\label{LK1}
\end{equation}
with
\begin{equation}
x= \left(2 \pi^2 \frac {k_B m_e}{ \hbar e}\right) p \mu T/B.  
\label{LK2}
\end{equation}
Here $p$ is the harmonic index of the specific oscillation, $\mu$ is the cyclotron effective mass in units of the electron mass, $T$ and $B$ are the temperature and magnetic field, and $m_e$ is the bare electron mass.\cite{lifshitz_theory_1958}  

%For simplicity, \eref{LK2} can be expressed numerically as 
%\begin{equation}
%x= 14.69~p \mu T/B
%\label{LK3}
%\end{equation}
%where $T$ is measured in Kelvin and $B$ in tesla.  
In \fref{fig3}(c) we fit our data to \eref{LK1}, including a small offset arising from imperfect background subtraction, and we find that $\mu=0.100 \pm 0.001$.  (The amplitude plotted is the amplitude of the peak at approximately 10.45\, T after background subtraction.)  
%Our fit of \eref{LK1} and \eref{LK2} to the decay of the Fourier amplitudes includes an additional two parameters corresponding to a scaling factor and an offset from zero.  This fit is shown in \fref{fig3}(b), which yields $\mu=0.037$.  
This mass is comparable to that observed in other topological systems, including Cd$_3$As$_2$,\cite{he_quantum_2014} TaAs,\cite{zhang_electron_2017} and Na$_3$Bi.\cite{kushwaha_bulk_2015}

The SdH oscillation amplitude also decays with $1/B$ due to the effect of finite carrier relaxation times according to the formula
\begin{equation}
R_D = \mathrm{exp}(-x T_D/T),
\label{dingle}
\end{equation}
where $T_D$ is called the Dingle temperature, and is inversely related to the relaxation time
\begin{equation}
\tau_D=\frac{\hbar}{2\pi k_B} \frac{1}{T_D}.
\label{dingle2}
\end{equation}
\cite{shoenberg_magnetic_1984}

Combining the $28\pm1\ \mathrm{T}$ oscillation frequency in \fref{fig3}(a) with the effective mass from \fref{fig3}(b) allows us to compute the various values in \tref{qo}.  Our reported oscillation frequency represents a Fermi-surface cross-sectional area of only 0.1\% of the perpendicular Brillouin zone area $(2 \pi / a)^2=2.1$\AA$^{-2}$.

The so-called Landau level fan diagram is frequently employed to measure the Berry's phase in suspected topological systems.  Conductivity minima in SdH oscillations (\fref{fig3}(b)) are plotted against their associated Landau level index.  The $x$-intercept of a linear fit to this data identifies the phase of the oscillations, and, in principle, the Berry's phase accumulated around one cyclotron orbit, according to the relation
\begin{equation}
\label{LL}
\Gamma = \frac{1}{2} - \frac{\phi_B}{2 \pi} + \delta,
\end{equation}
in which $\Gamma$ is the $x$-intercept, $\phi_B$ the Berry's phase, and $\delta$ a number less than 1/8 in magnitude.\cite{sergelius_berry_2016,lukyanchuk_dirac_2006,cao_landau_2015,lukyanchuk_phase_2004,pariari_probing_2015}
Some authors suggest that $\delta$ is related to Fermi surface curvature: a smooth, 2D cylinder yields $\delta=0$, but a ``corrugated 3D Fermi surface'' merits $|\delta|=1/8$.\cite{lukyanchuk_phase_2004}  The angular dependence of the oscillation frequency (\fref{fig3}(e) and discussed below) indicates our system is close to the 2D cylinder case, suggesting $\delta \approx 0$.  In this case, a topologically nontrivial Berry's phase of $\pi$ yields $\Gamma=0$, while a trivial Berry's phase of 0 yields $\Gamma=1/2$.  The actual value we measure in \fref{fig3}(d) is $\Gamma=\num{8e-4} \pm \num{1.5e-3}$.  This value indicates that the Fermi surface pocket we probe with SdH oscillations is consistent with having a topological origin.

\begin{table}
\caption{\label{qo}Electronic properties from Shubnikov-de Haas oscillations. Here $F$ is the SdH frequency, $k_F$ is the Fermi wavenumber, $m_c$ is the cyclotron mass, $n_{SdH}$ is the carrier density, and $T_D$ is the Dingle temperature.}
\begin{ruledtabular}
\begin{tabular}{lll}
\textbf{Parameter} & \textbf{Value}\\
$F$ (T) & $28\pm1$\\
%Extremal orbit area (\AA$^{-2}$) & $2.8 \times 10^{-3}$\\
$k_F$ (nm$^{-1}$) & $0.29 \pm 0.01$\\
%$E_F$ (meV) & $91\pm3$ &\\
$m_c$ (m$_\mathrm{e}$) & $0.100 \pm 0.001$\\
$n_{SdH}$ (cm$^{-3}$)& $(5.0 \pm 0.1)\times 10^{19}$\\
$T_D$\,(K) & $23\pm2$\\
%Dingle lifetime, $\tau_D$\,(s) & $(3.0\pm0.5)\times 10^{-14}$\\
\end{tabular}
\end{ruledtabular}
\end{table}

By measuring SdH oscillations at different angles of the applied field, we are able to use the angular variation of the oscillation frequency to map the shape of the Fermi surface.  The angular dependence of the SdH frequency is shown in \fref{fig3}(e), with a 5\textdegree\ sample positioning error subtracted.  (After about 75\textdegree\ the oscillation amplitude has decreased in amplitude beyond detection.)  Note that for a spherical Fermi surface, the data points in \fref{fig3}(e) would fall on a circle.  Instead, they more closely follow a $1/\mathrm{cos}(\theta)$ dependence consistent with a two-dimensional or strongly anisotropic three-dimensional Fermi surface.

\section{DISCUSSION}
The angular dependence of the Shubnikov-de Haas oscillation frequency is consistent with a small 2D Fermi surface.  This result is commensurate with our ARPES measurements, which show small 2D pockets along the $(0,\pi)$ and $(\pi, \pi)$ directions in the Brillouin zone.  Moreover, both ARPES and the Landau level fan diagram suggest that these pockets originate from a Dirac band crossing.  \yms\ may therefore be a new topological material.

ARPES and DFT calculations indicate that three Fermi surface pockets contribute carriers: (1) Small 3D hole-like pockets ($\alpha$), with a large effective mass, (2) 2D electron-like cylinders ($\beta$), with a small effective mass, and (3) small 2D hole-like elliptic cylinders ($\gamma$), also with a small effective mass.

Using the SdH oscillation data in \tref{qo}, we can place an upper bound on the low temperature resistivity with a simple Drude approximation for a single pocket.  This value is an upper bound both because the Dingle lifetime is sensitive to small-angle scattering (which only weakly affects the resistance) and because it ignores all other Fermi pockets.\cite{shoenberg_magnetic_1984}  Assuming a cylindrical Fermi surface, this bound is $133\pm13$\ $\mu \Omega\mathrm{cm}$, with the uncertainty primarily arising from the determination of the Dingle lifetime. Our observed resistivity at low temperature is $160\pm16$\ $\mu \Omega\mathrm{cm}$, which nearly coincides with this upper bound.  (The uncertainty in this resistivity originates from the measurement of the sample thickness.)  This observation would imply that \yms\,is in an unusual case whereby the quantum lifetime matches the transport lifetime. There are two plausible explanations for this.  The first is that, the quantum oscillations we observe do not originate from bulk bands but in fact originate from high mobility surface states that carry most of the current. However, our preliminary studies show that there is a weak thickness dependence to the resistivity, suggesting surface states do not contribute significantly to the current. 

The second scenario is that, although the quantum (Dingle) lifetime tends to underestimate the transport (Drude) lifetime due to its sensitivity to small-angle scattering, the Fermi surface in this material is quite small and so small-angle scattering may be catastrophic, hence efficiently dissipating momentum.  The remainder of the carriers must then have very low mobility, and contribute relatively little to conduction having been ``shorted out'' by the mobile carriers observed in quantum oscillations. In this scenario, the SdH oscillations must originate from either the 2D $\beta$ or $\gamma$ pockets since the $\alpha$ pockets can be ruled out because they are 3D.  The Hall effect (\fref{fig2}(c)) indicates that hole-like carriers are more numerous and more mobile, suggesting the $\gamma$ pocket is the origin of these oscillations.  On the other hand, the balance of carriers and mobilities may conspire to achieve an identical Hall signal such that the $\beta$ pockets cannot be ruled out. Nevertheless, the balance of evidence suggests this second scenario most likely explains our data.

Finally, we emphasize that our ARPES data cannot be reconciled with our DFT calculations without antiferromagnetic order on the Mn sites. Thus, although we do not have any direct thermodynamic evidence for an antiferromagnetic phase transition,  the close agreement between the calculations and experiment strongly suggests the presence of magnetic order.  It may be that the thermodynamic anomaly is weak and cannot be resolved, or occurs at temperatures beyond those measured here.  Compounds with identical Mn sublattices, including EuMnBi$_2$ ($T_N\approx$\ 310\,K)\cite{may_effect_2014} and SrMnBi$_2$ ($T_N\approx$\ 290\,K),\cite{park_anisotropic_2011} exhibit antiferromagnetic ordering near room temperature.  

\section{Conclusion}
We have performed a comprehensive study of the candidate topological semimetal \yms\ using electrical transport and ARPES measurements, all of which can be reconciled with our detailed DFT calculations.  Shubnikov-de Haas oscillations reveal a 2D Fermi surface of Dirac origin, as predicted by DFT and consistent with ARPES measurements.  This work indicates that \yms\ is a possible new test bed for the study of topological states of matter in the presence of broken-symmetry order.

\section{Acknowledgments}
R. K. is supported by the National Science Foundation (NSF) Graduate Research Fellowship under Grant No. DGE-1106400. S. J., C. J., J. G. A., and much of this work received support from the Gordon and Betty Moore Foundation under Grant No. GBMF4374.  S. J., S. M. G., J. D. D. and J. B. N. were supported by the Director, Office of Science, Office of Basic Energy Sciences, Materials Sciences and Engineering Division, of the U.S. Department of Energy under Contract No. DE-AC02-05-CH11231.  Computational resources provided in part by the Molecular Foundry were supported by the Office of Science, Office of Basic Energy Sciences, of the U.S. Department of Energy, also under Contract No. DE-AC02-05-CH11231. S. M. G. acknowledges financial support by the Swiss National Science Foundation Early Postdoctoral Mobility Program.  K. A. B. and J. Y. C. acknowledge partial support from NSF DMR-1360863.  This paper also benefited from Dr. Nicholas P. Breznay's constructive feedback.

%\bibliography{Kealhofer_YbMnSb2.bib}

\end{document}